\pdfoutput=1
\documentclass[runningheads]{llncs}
\PassOptionsToPackage{usenames,dvipsnames,svgnames,table}{xcolor}

\usepackage[frozencache=true,cachedir=.//,newfloat,langlinenos=true]{minted}
\usepackage{plantuml}
\usepackage{float}

\usepackage{algorithm}
\usepackage{algpseudocode}
\usepackage{multirow}
\usepackage[colorinlistoftodos,prependcaption,textsize=tiny]{todonotes}
\usepackage[normalem]{ulem}

\usepackage{graphicx}                   
\input{macros}

\begin{document}

\title{A Runtime Environment for Contract Automata\thanks{This version of the contribution has been accepted for publication, after peer review (when applicable)
but is not the Version of Record and does not reflect post-acceptance improvements, or any corrections.
The Version of Record is available online at: https://doi.org/10.1007/978-3-031-27481-7\_31.
Use of this Accepted Version is subject to the publisher’s Accepted Manuscript terms of use
https://www.springernature.com/gp/open-research/policies/accepted-manuscript-terms}}
\titlerunning{Contract Automata Runtime Environment}

\author{Davide Basile\,\textsuperscript\Letter\,\orcidID{0000-0002-7196-6609} \and Maurice H. ter Beek\orcidID{0000-0002-2930-6367} }
\authorrunning{D. Basile and M.H. ter Beek }

\institute{Formal Methods and Tools Lab\\ 
ISTI--CNR, Pisa, Italy\\
\email{\{davide.basile,maurice.terbeek\}@isti.cnr.it}
}

\maketitle

\sloppy
\vspace*{-1.25\baselineskip}

\begin{abstract}
Contract automata have been introduced for specifying  applications through behavioural contracts 
and for synthesising their orchestrations as finite state automata.  
This paper addresses the realisation of applications from contract automata specifications. 
We present {\tt CARE}, a new runtime environment to coordinate services implementing contracts that guarantees the adherence of the implementation to its contract. 
We discuss how {\tt CARE} can be adopted to realise contract-based applications, its formal guarantees, and we identify the responsibilities of the involved business actors. 
Experiments show the benefits of adopting {\tt CARE} with respect to manual implementations.
\end{abstract}

\section{Introduction}\label{sec:introduction}


%
From a recent survey in the transport domain~\cite{FB22}, it has emerged that the majority of studies on formal methods  propose  specification languages, models, and their verification, 
whereas fewer focus on how to  derive the finalised software from the verified specification also showing the adherence of the implementation to its specification.  
The authors of~\cite{DBLP:journals/scp/KouzapasDPG18} state that 
 these interaction specifications {``are not yet a feature of standard mainstream programming languages, so software developers are not able to benefit from them''}. 
 %
%
%
%
In this paper, we investigate the connection between a behavioural specification and its implementation, and we provide a possible realisation of those aspects abstracted in a specification. 

Contract automata are a dialect of finite state automata used to formally specify the behaviour of service contracts 
in terms of offers and requests~\cite{BDF16}. 
A composition of contracts is in  {\it agreement} when all requests are 
matched by corresponding offers of other contracts.
A composition can be refined to one in agreement using 
 the orchestration synthesis algorithm~\cite{BBDLFGD20,BBP20}, 
 a variation of the synthesis algorithm from 
 supervisory control theory~\cite{RW87}.
Previously, in~\cite{BasileB21}, a library  called~\texttt{CATLib}~\cite{BB22OSP}  implementing the operations on contract automata (e.g., composition, synthesis) was presented. 
A front-end of~\texttt{CATLib}  for graphically editing
and operating on contracts is also available~\cite{CATAPPurl}, called \texttt{CAT\_App}.
The orchestrator is abstracted away in contract automata and until now no examples of concrete implementations were provided in which services implement contract automata specifications.

Whilst \texttt{CATLib} and \texttt{CAT\_App} are used to \emph{specify} applications as contract automata, 
in this paper we tackle the problem of \emph{implementing} applications that have been specified via contract automata.  
We introduce {\tt CARE}~\cite{CAREurl}, a newly developed software that provides a runtime environment to coordinate the {\tt CARE} services that implement the contracts of the synthesised orchestration. 
Thus, {\tt CARE} advances the state-of-the-art of the research on contract automata and behavioural contracts by detailing how specifications through contract automata can be connected with service-based  applications. With {\tt CARE}, the low-level interactions that are abstracted in contract automata orchestrations are now explicated. 
We discuss how {\tt CARE} can promote a separation of concerns among different actors that together cooperate  to realise contract-based applications, and among developers and designers of services. 
The proposed framework is exercised on two examples, showcasing the usage of {\tt CARE}. 
Experiments show a neat improvement in terms of decreased complexity of the software when comparing the  implementations of the examples exploiting {\tt CARE}  with those that manually implement the low-level interactions among services without relying on {\tt CARE}. 

\paragraph{\bf Related Work}
\label{sect:relatedwork}
Other approaches to connect implementations with behavioural types (e.g., behavioural contracts, session types)  are surveyed in~\cite{DBLP:journals/ftpl/AnconaBB0CDGGGH16,Gay20171}. 
Our approach is closer to~\cite{DBLP:journals/scp/KouzapasDPG18,DBLP:journals/corr/abs-2009-08769}, where behavioural types are expressed as finite state automata of {\tt Mungo}, called {\it typestates}~\cite{DBLP:journals/tse/StromY86}. 
The toolchain of {\tt Mungo} and {\tt StMungo} is proposed to implement behavioural types specifications. 
Similarly to {\tt CARE}, in {\tt Mungo} finite state automata are used as behaviour assigned to Java classes (one automaton per class), where transition labels correspond to methods of the classes.  
A tool similar to {\tt Mungo} is {\tt JaTyC} (Java Typestate Checker)~\cite{BACCHIANI2022102844}.

An Eclipse plugin called {\tt Diogenes}~\cite{DBLP:conf/forte/AtzeiB16}  allows to write specifications of services as behavioural contracts using a domain specific language, verify them, and generate skeletal Java programs to be refined using the Java RESTful Web service middleware for contract-oriented computing presented in~\cite{DBLP:conf/facs2/BartolettiCMPP15}. 
Both {\tt Diogenes} and {\tt StMungo}  generate skeletal Java programs from contract compositions or multi-party session types, respectively, whereas {\tt CARE} allows to adapt already existing  components to realise a new application in a bottom-up approach, fostering adaptability and reusability of services.

{\tt CARE} adopts a  {\it correct-by-design} approach to implement a specification with formal guarantees. 
The complementary approach infers a behavioural type from an  implementation, where guarantees hold if the typing succeeds.
An algorithm to infer  a form of behavioural types from programs  with assertions is discussed in~\cite{DBLP:conf/sac/VasconcelosR17}, where programs are written in {\tt Mool} (Mini object-oriented language), a simple Java-like language incorporating behavioural types.  
The inference of  behavioural types from {\tt Go} programs is studied in~\cite{DBLP:conf/icse/LangeNTY18}. 
 {\tt Go} is a language supporting synchronisations 
on  channels  inspired by process algebraic formalisms like CSP and CCS~\cite{DBLP:conf/wcre/DilleyL19}. 
The inference of behavioural types is thus facilitated by the chosen languages, whilst extracting them from unconstrained Java programs is still a challenge~\cite{DBLP:conf/fmics/RubbensLH21}.  
{\tt CATLib} supports compositions of communicating machines, the formalism of behavioural types used in~\cite{DBLP:conf/icse/LangeNTY18}, thus it could be used   to suggest amendments to the original {\tt Go} programs by exploiting its synthesis algorithms.

Finally, the approach proposed by {\tt CARE} shares aspects with the synthesis/verification of runtime  monitors~\cite{DBLP:journals/sttt/AcetoCFI21,DBLP:journals/fmsd/SanchezSABBCFFK19,DBLP:conf/atva/AzzopardiPS21}, and 
 is similar to the {\it automated composition} problem  studied in~\cite{DBLP:journals/deds/AtamporeDR19,DBLP:conf/smc/BaratiS15,DBLP:journals/access/Farhat18,7473906}, to which {\tt CARE} and {\tt CATLib} 
offer both a runtime engine and tailored novel synthesis algorithms. 


\paragraph{\bf Outline}
We provide some  
background on contract automata in Section~\ref{sect:background}.  
Section~\ref{sect:design} details the design of {\tt CARE}.  
The formal guarantees offered by {\tt CARE} are detailed in Section~\ref{sect:correctness}, whilst Section~\ref{sect:methodology} discusses how {\tt CARE} can be adopted for building applications specified via contract automata and the separation of concerns. 
Section~\ref{sect:usage} contains two examples and 
an evaluation of the benefits of our contribution. 
Finally, we conclude and discuss future work in Section~\ref{sec:conclusion}.

\section{Modal Service Contract Automata}\label{sect:background}

We provide 
background on contract automata and their synthesis operation.

A Contract Automaton (CA) models either a single service or a multi-party composition of services performing actions. 
Figure~\ref{fig:contracts} depicts some examples of CA.
The number of services of a CA is called its \emph{rank}. 
When $\textit{rank}=1$, the contract is called a \emph{principal} (i.e., a single service). 
For example, the leftmost and rightmost automata in Figure~\ref{fig:contracts} are principals, while the automaton in the middle has $\textit{rank}=2$.
Labels of CA are vectors of atomic elements called \emph{actions}.
Actions are either \emph{requests\/} (prefixed by~{\tt ?}),  \emph{offers\/} (prefixed by~{\tt !}), or \emph{idle\/} 
(denoted with a distinguished symbol~{\tt -}). 
Requests and offers belong to the (pairwise disjoint) sets $\Rset$ and $\Oset$, respectively. 
The states of CA are vectors of atomic elements called basic states. 
Labels are restricted to be 
\emph{requests\/}, \emph{offers\/}, or \emph{matches\/} where, respectively,
there is either a single request action, a single offer action, or a single pair of request and offer actions that match, and all other actions are idle. 
The length of the vectors of states and labels is equal to the rank of the CA.

For example, the label {\tt [!euro, ?euro]} is a match where the request action {\tt ?euro} is matched by the offer action {\tt !euro}. 
Note the difference between a request label (e.g., {\tt [?coffee,~-]}) and a request action (e.g., {\tt ?coffee}). 
A transition may also be called a request, offer, or match according to its label.

The goal of each service is to reach an accepting (\emph{final}) state such that all its request (and possibly offer) actions are matched.
In~\cite{BBDLFGD20}, CA were equipped with \emph{modalities}, i.e., \emph{necessary\/}~($\Box$) and \emph{permitted\/}~($\Permitted$) transitions, respectively. 
Permitted transitions are controllable, whereas necessary transitions can be uncontrollable or semi-controllable. 
The resulting formalism is called \emph{Modal Service Contract Automata} (MSCA). 
In the following definition, given a vector $\vec a$, its $i$th element is denoted by $\ithel {\vec a} i$.

\begin{definition}[MSCA]\label{def:contract}
Given a finite set of basic states $\mathcal{Q} = \{q_1,q_2, \ldots \}$, an 
MSCA $\mathcal{A}$ of $\textit{rank}=n$ is a tuple $\langle Q, \vec{q_0}, A^r, A^o, T, F \rangle$, 
with set of states $Q=Q_1 \times \ldots \times Q_n \subseteq \mathcal{Q}^n$, 
initial state $\vec{q_0} \in Q$,
%
%
set of requests $A^{r} \subseteq \Rset$,
set of offers $A^{o} \subseteq \Oset$, set of final states $F \subseteq Q$,
set of transitions $T \subseteq Q \times A  \times Q$, where 
$A\subseteq(A^r \cup \offerset \cup \{{\texttt-} \})^n$,  
partitioned into \emph{permitted\/} transitions $T^\Diamond$ and \emph{necessary\/} transitions $T^\Box$, 
such that: 
(i)~given $t= (\vec{q},\vec{a},\vec{q}\,') \in T$, $\vec{a}$ is either a request, or an offer, or a match;
and (ii)~$\forall i \in 1\ldots n,\ \ithel{\vec{a}} i={\texttt -}$ implies $\ithel{\vec{q}} i=\ithel{\vec{q}\,'} i$.
\end{definition}

Composition of services is rendered through the composition of their MSCA models by means of the \emph{composition operator\/} $\otimes$, which is a variant  of a synchronous product. 
This operator basically interleaves or matches the transitions of the component MSCA, but whenever two component MSCA are enabled to execute their respective request/offer, then the match is forced to happen.
Moreover, a match involving a necessary transition of an operand is itself necessary.
The rank of the composed MSCA is the sum of the ranks of its operands. The vectors of states and actions of the composed MSCA  are built from the vectors of states and actions of the component MSCA, respectively.
Typically, in a composition of MSCA various properties are analysed. We are especially interested in \emph{agreement\/}. In a contract that is in agreement, all requests are matched, i.e., transitions are only labelled with offers or matches. 


We recall the specification of the abstract synthesis algorithm of CA from~\cite{BBP20}.
The synthesis of a controller, an orchestration, and a choreography of CA are all different special cases of this abstract synthesis algorithm, formalised in~\cite{BBP20} and implemented in \texttt{CATLib}~\cite{BasileB21}.
This  algorithm is a fixed-point computation where at each iteration the set of transitions of the automaton is refined (using pruning predicate~$\phi_p$) and a set of forbidden states $R$ 
is computed (using forbidden predicate~$\phi_f$).
The synthesis is parametric with respect to these two predicates, which provide information on when a transition has to be pruned from the synthesised automaton and when a state has to be deemed forbidden, respectively. 
We refer to MSCA as the set of (MS)CA, where the set of states is denoted by $Q$ and the set of transitions by $T$ (with $T^\Box$ denoting the set of necessary transitions). 
For an automaton $\mathcal A$, the predicate $Dangling(\mathcal A)$ contains those states that are not reachable from the initial state or that cannot reach any final state.

\begin{definition}[abstract synthesis~\cite{BBP20}]\label{def:abstractsynthesis}
Let $\mathcal{A}$ be an MSCA, $\mathcal{K}_{0} = \mathcal A$, and $R_{0} = \textit{Dangling\/}(\mathcal{K}_{0})$. 
Given two predicates $\phi_p, \phi_f: T \times \textit{MSCA} \times Q \rightarrow \mathbb{B}$,
let the \emph{abstract synthesis function\/} $f_{(\phi_p, \phi_f)}: \textit{MSCA} \times  2^Q \rightarrow \textit{MSCA} \times  2^Q$ be defined as: 
\begin{center}
\def\arraystretch{1.25}
\begin{tabular}{c@{\hskip 0.03in}r@{\hskip 0.03in}c@{\hskip 0.03in}l}
\multicolumn{4}{l}{
$f_{(\phi_p, \phi_f)}(\mathcal{K}_{i-1},R_{i-1}) = (\mathcal{K}_{i},R_{i})$, with}\\
& $T_{\mathcal{K}_{i}}$ & = & $T_{\mathcal{K}_{i-1}} -  
\{\, t \in T_{\mathcal K_{i-1}} \mid   
                                        \phi_p(t, \mathcal K_{i-1}, R_{i-1}) = \textit{true\/}
                                        \,\}$\\
& $R_{i}$ & = & $R_{i-1} \cup
\{\,\vec q \mid (\vec q \TRANS{}) = t \in T_{\mathcal A}^\Box\,,\  \phi_f(t, \mathcal K_{i-1}, R_{i-1}) = \textit{true\/} \,\} \cup \textit{Dangling\/}(\mathcal{K}_{i})$
\end{tabular}
\end{center}
\end{definition}
Subsequently, the abstract controller is defined 
as the least fixed point  of $f_{(\phi_p, \phi_f)}$ (cf.~\cite[Theorem 5.2]{BBP20}). 
The synthesised orchestration guarantees the reachability of final states, the agreement property (i.e., all requests are matched) and that all reachable necessary requests are not pruned (i.e., controllability).
%

\paragraph{\bf Tooling}
CA and their functionalities are implemented in a software artefact, called Contract Automata Library (\texttt{CATLib}), whose development is active~\cite{BB22OSP}.
This software artefact is a by-product of our scientific research on behavioural contracts and implements results that have previously been formally specified in 
several publications (cf., e.g.,~\cite{BBDLFGD20,BBP20,BDF16}). 
 Scalability features offered by {\tt CATLib} include a bounded on-the-fly state-space generation optimised with pruning of redundant transitions and parallel streams computations. 
 The software is open source~\cite{BB22OSP}, it has been developed using principles of model-based software engineering~\cite{BasileB21} and it has been extensively validated using various testing and analysis tools to increase the confidence on the reliability of the library~\cite{BB22OSP}.


\vspace*{-0.3cm}
\section{{\tt CARE} Design}\label{sect:design}

\begin{figure}[t]
    \centering
    \includegraphics[width=.95\columnwidth]{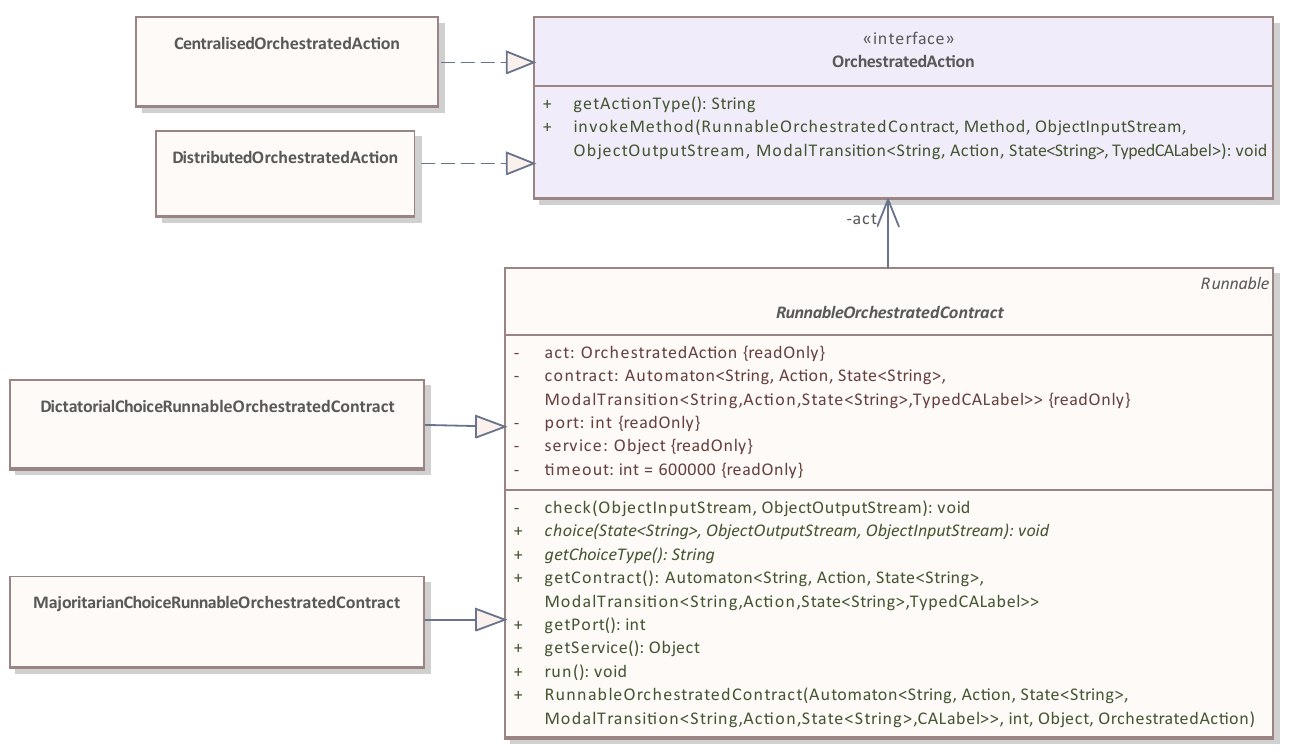}
    \caption{The class diagram for the orchestrated services;
the methods of the derived classes are visible in their super-class/interface as abstract methods (in italic)}
    \label{fig:orchestrated}
\end{figure}

We start by discussing the design of {\tt CARE}. 
This software is organised into classes for the orchestrated services (cf.\ Figure~\ref{fig:orchestrated}) and classes for the orchestrator.

In Figure~\ref{fig:orchestrated}, \texttt{RunnableOrchestratedContract} is an abstract class that implements an executable wrapper responsible for the composition of  the specification of a service (instance variable {\tt contract} storing a contract automaton) with its implementation (instance variable {\tt service} implementing the service).
\texttt{RunnableOrchestratedContract} implements a service that is always listening and spawns  a parallel process when entering an orchestration. 
During an orchestration, it receives action commands from the orchestrator or from other services, and it invokes the corresponding action method (by means of the instance variable {\tt act} of type {\tt OrchestratedAction}).

The realisation of an orchestration is abstracted away in contract automata. 
Crucially, offers and requests of contracts are an abstraction of low-level messages sent between the services and the orchestrator to realise them.
{\tt CARE}  exploits the abstractions provided by Java to allow its specialisation according to different implementation choices, using abstractions of object-oriented design, as showed in Figure~\ref{fig:orchestrated}. 
Two  aspects to implement are choices and termination (through the abstract method {\tt choice}). 
{\tt CARE} is equipped with default implementations, but can be extended (by implementing the relative interfaces and abstract methods) to include other options, other than the default ones.
Currently, a so-called `dictatorial' choice (i.e., an internal choice of the orchestrator, external for the services) and a so-called `majoritarian' choice (services vote 
and the majority wins) are two implemented options. 
{\tt MajoritarianChoiceRunnableOrchestratedContract} and {\tt DictatorialChoiceRunnableOrchestratedContract} are the two classes specialising {\tt RunnableOrchestratedContract} according to how the choice is handled and implementing the abstract methods.
{\tt CARE} also provides default implementations for  the low-level message exchanges. 
Currently, the two available options are the `centralised' action, where the orchestrator acts as a proxy, 
and the `distributed' action, where two  services matching their actions directly interact with each other  once the orchestrator has made them aware of a matching partner and its address/port. 
Accordingly, each {\tt RunnableOrchestratedContract} has an {\tt OrchestratedAction} (instance variable {\tt act}) used to implement the corresponding actions that can be either distributed or centralised according to the current implementation. 
%

The abstract class \texttt{RunnableOrchestration} (which is not displayed in Figure~\ref{fig:orchestrated}) implements a special service that reads the synthesised orchestration (stored in the instance variable {\tt contract}) and orchestrates the \texttt{RunnableOrchestratedContract} to realise the overall application.  
%
Similarly to the case of the orchestrated contract, also the 
orchestrator is specialised according to either a dictatorial or a majoritarian implementation of the abstract method {\tt choice}.
Moreover, an {\tt OrchestratorAction} instance variable is used to implement each action of the orchestration, either centralised or distributed, thus matching the corresponding actions of the orchestrated services.

Finally, 
 the class \texttt{ContractViolationException} implements an exception raised in case an invocation of the orchestrator is not allowed by the orchestrated contract or if that contract is not fulfilled. 
 When thrown, the exception stores the remote host that violates the contract. This guarantees 
 the accountability in case of a contract violation. 
Each label of a contract automaton is extended using {\tt CARE} with the 
information on the types of parameters and returned values from the corresponding method implementing the corresponding action. 
These typed labels are implemented into the class {\tt TypedCALabel}, extending a {\tt CALabel} of {\tt CATLib}.
This class also overrides the matching between requests and offers to also take into account  
  their types: the returned value of the request must be of a super-class of the parameter class 
of the offer and vice versa. 
This guarantees that no {\tt ClassCastException} will ever be 
raised when invoking the actions.
Note that the signature of each action declared by the interface is not fixed, so
other types can be used (e.g., \texttt{JSon} values).

\begin{wrapfigure}[10]{r}{0.3\textwidth}
\vspace{-0.7cm}
\centering\includegraphics[width=0.3\textwidth]{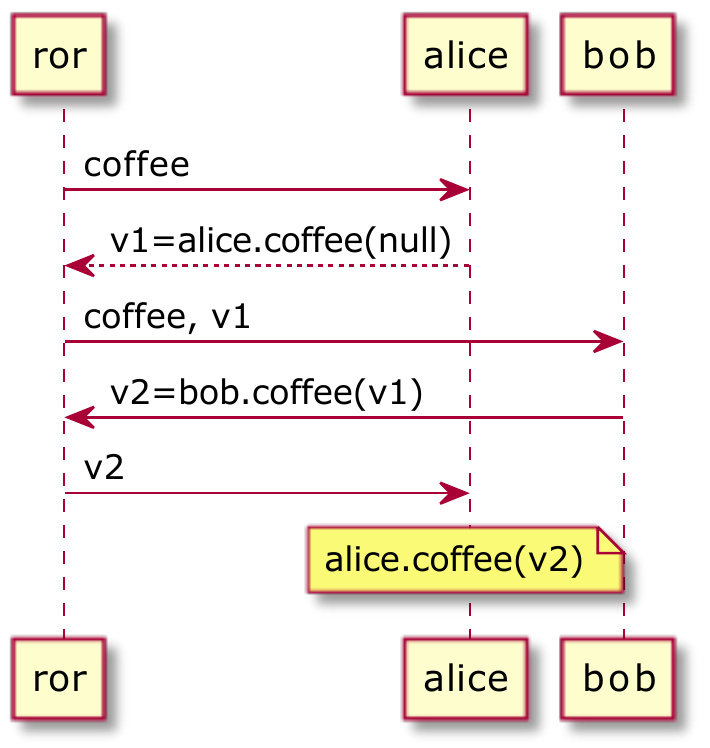}
\end{wrapfigure}
We briefly detail the centralised implementation of a match label  in {\tt CARE}. We will use the match  \texttt{[?coffee,!coffee]} from the example in Section~\ref{sect:usage}, in which \texttt{Alice} is requesting a \texttt{coffee} and \texttt{Bob} is offering a \texttt{coffee}.
%
%
%
%
The method \texttt{coffee} of \texttt{Alice}  is invoked twice: firstly, passing no argument, it generates an {\tt Integer} value (e.g., the amount of sugar) that is passed (by the orchestrator {\tt ror}) as argument to the method \texttt{coffee} of \texttt{Bob}, which in turn produces a {\tt String} value that is eventually passed  as  argument to the method \texttt{coffee} of \texttt{Alice}, thus fulfilling the {\tt coffee} request.
%

%
%

\vspace{-0.3cm}
\section{Formal Guarantees}\label{sect:correctness}
We now discuss the formal guarantees of correctness and the adherence of the implementation to the specification brought by the usage of {\tt CARE}. 
To begin with, to guarantee that an orchestration  is  {\it correct-by-design}, 
the contract automata operators of composition and orchestration synthesis are used, exploiting the theoretical results on contract automata (cf.\ Section~\ref{sect:background}). 
More concretely, these operations are performed in the constructor of a {\tt RunnableOrchestration} using {\tt CATLib}. 
As discussed in Section~\ref{sect:background}, the synthesised orchestration ensures properties such as absence of deadlocks, matching of all requests of contracts with corresponding offers of other contracts, and reachability of final states.

After a well-behaving orchestration has been synthesised, it is important to ensure that the low-level implementations of the distributed services interacting with each other will adhere to the operations prescribed by the orchestration synthesised from their contracts.
This task is addressed by using Algorithm~\ref{alg:orchestration} and Algorithm~\ref{alg:service}, both implemented in {\tt CARE}, reproduced below in pseudo-code.

\medskip
\setlength{\intextsep}{0pt} 
\noindent
\begin{minipage}{.495\columnwidth}
\vspace{-0.84cm}
\begin{algorithm}[H]
\scriptsize
\caption{Orchestration}\label{alg:orchestration}
\begin{algorithmic}
\Require non-empty orchestration automaton
\Ensure no exception is thrown
\State \textit{init Sockets} \Comment{connect to the services}
\State \textit{cs} $\gets$ \textit{initialState}   \Comment{current state}
\While{\textit{true}}
    \State \textit{fws} $\gets$ \textit{forwardStar(cs)}
\If{\textit{empty(fws)} $\And$ \textit{notFinal(cs)}}
    \State  {\bf throw} Exception
\EndIf
\State \textit{choice} $\gets$ \textit{choice()} \Comment{interact with services}
\If{\textit{choice} $==$ stop $\And$ \textit{final(cs)}}
    \State \textbf{return}
\EndIf
\State \textit{tr} $\gets$ \textit{select(fws,choice)}
\If{$tr$ not in agreement} \State  {\bf throw} Exception
 \EndIf
 \State \textit{doAction(tr)} \Comment{interact with services}
 \State \textit{cs} $\gets$ \textit{targetState(tr)}
\EndWhile
\end{algorithmic}
\end{algorithm}
\end{minipage}\hfill\vline\hfill
\begin{minipage}{.495\columnwidth}
  \begin{algorithm}[H]
\scriptsize
\caption{Service Thread}\label{alg:service}
\begin{algorithmic}
\Require connected to the orchestrator
\State \textit{init Socket}  \Comment{set socket timeout}
\State \textit{cs} $\gets$ \textit{initialState}   \Comment{current state}
\While{true}
\State \textit{act} $\gets$ \textit{receive(socket)}
\If{\textit{stop(act)}}
\If{\textit{final(cs)}}
\State {\bf return}
\Else \ {\bf throw} ContractViolationException
\EndIf
\EndIf
\If{$choice(act)$}
\State \textit{performChoice()} \Comment{interact with or-}
\State {\bf continue} \Comment{chestration}
\EndIf
\State \textit{tr} $\gets$ \textit{select(forwardStar(cs),act)}
\If{no valid action}
\State {\bf throw} ContractViolationException
\Else 
\State \textit{invokeMethod(tr)}
\EndIf
\State \textit{cs} $\gets$ \textit{targetState(tr)}
\EndWhile
\end{algorithmic}
\end{algorithm}
\end{minipage}
\medskip

Algorithm~\ref{alg:orchestration} illustrates the main operations performed during an  orchestration. The algorithm requires that a correct and non-empty orchestration has been synthesised. This requirement is necessary to ensure that no exceptions will be thrown at runtime.  
Initially, the orchestrator connects to the services (their ports and addresses are stored during instantiation). 
The current state of the execution is set to  the initial state. 
Subsequently, a loop is executed in continuation. 
Inside the loop, one of the transitions is selected from  the set of outgoing transitions (i.e., the forward star) of the current state, using the implementation of the abstract method {\tt choice}. Here, if there is a deadlock (no outgoing transitions and the current state is not final), an exception is thrown. 
After that, if the current state is final, but there are also outgoing transitions, then the choice can be to stop or to continue; otherwise, if the state is not final, then the choice can only be to select one of the outgoing transitions. 
If the selected transition of the orchestration does not satisfy agreement (i.e., its label is a request), then an exception is thrown. 
Otherwise, the action of the selected transition is executed using the implementation of the abstract method {\tt doAction} and the current state is updated to the target state of the transition.
As discussed in Section~\ref{sect:background}, if the orchestration automaton has been  synthesised using the  contract automata synthesis, then this formally guarantees that the described exceptions will never be thrown by the orchestrator. 

Algorithm~\ref{alg:service} summarises the execution of an orchestrated service following its contract. The service is multi-threaded and spawns a new thread each time a new request of connection is received. The algorithm depicts the operations performed by a spawned thread. Similarly to the orchestration, there is an initialisation of the socket, and the current state is set to the initial state of the contract. After that, a continuous loop is executed. 
Firstly, an action is received from the orchestrator. 
If the choice is of terminating and the contract is in a final state, then the service terminates successfully; otherwise, if the state is not final, an exception is thrown. 
If the orchestrator requires to make a choice, then the implementation of the abstract method {\tt choice} is called to perform a choice (possibly interacting with the orchestrator). 
Otherwise, the orchestrator is requiring to perform an action. 
In this case, the prescribed action is selected from the outgoing transitions of the current state of the service contract. 
If there is no such action, then a contract violation exception is thrown since the orchestrator is requiring to perform an operation not prescribed by the contract. 
Otherwise,  the method of the service that is paired with the corresponding action of the contract is invoked.  
These steps ensure that the low-level implementation of the actions of the services are correctly executed according to the actions prescribed by the orchestration synthesised from the composition of contracts.
Finally, the current state is set to the target state of the contract and the loop is repeated. 
Similarly to the orchestration case, if the orchestrator is executing a correctly synthesised orchestration, then the services will never throw  any such exception. 
Indeed, this would be a contradiction to the formal results discussed in Section~\ref{sect:background}.

\begin{figure}[t]
    \includegraphics[width=\columnwidth]{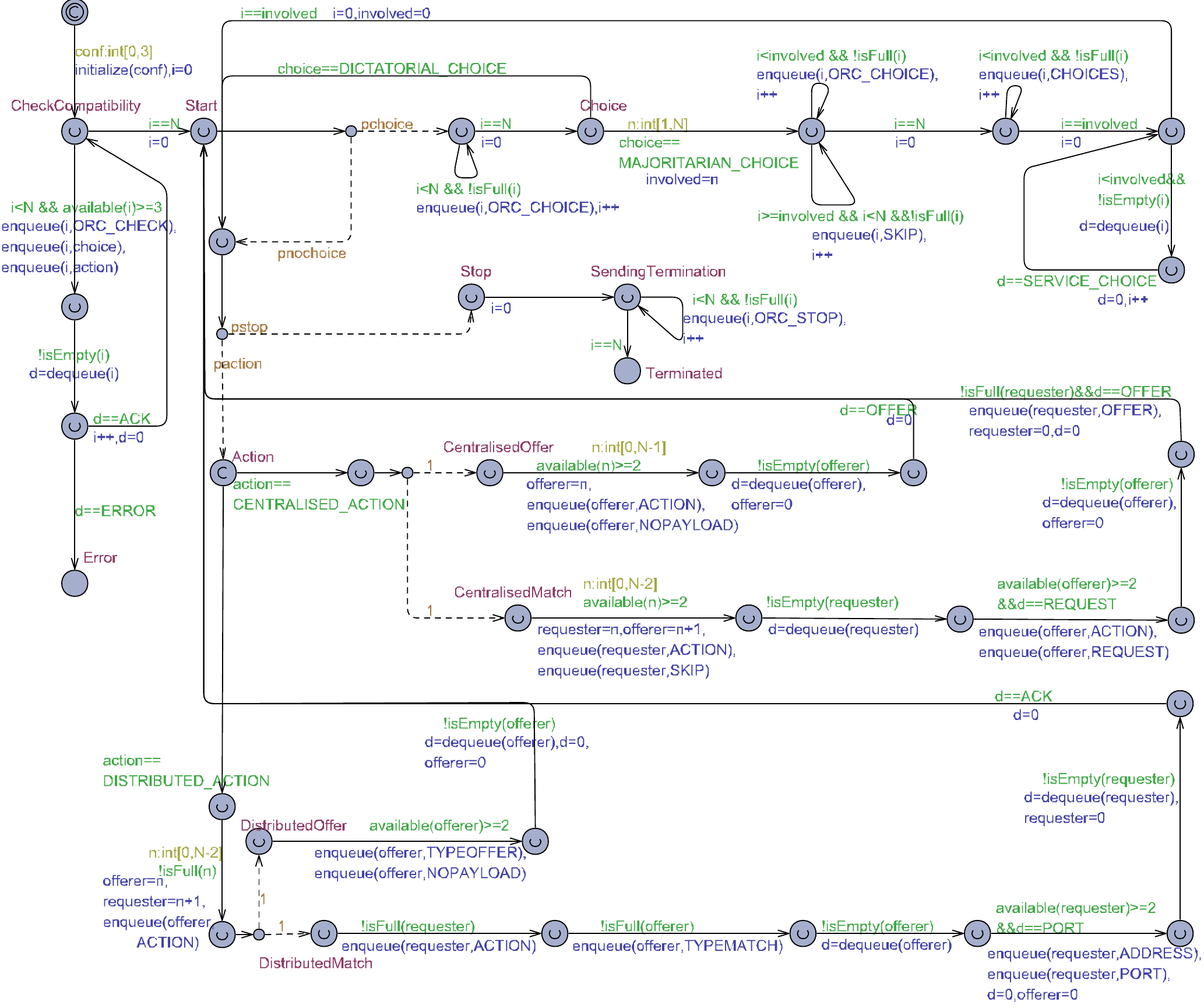}
    \vspace*{-0.5cm}
    \caption{\label{fig:runnableorchestration}The {\tt RunnableOrchestration} {\sc Uppaal} model}
\end{figure}

\paragraph{\bf Interaction Correctness} 
As stated above, the execution of an \emph{action} or a \emph{choice}  is  abstracted in {\tt CARE}.  Two  implementations are currently available for both actions and choices, and the framework is extensible.
We now summarise the formal verification of the TCP/IP sockets interactions performed by the available implementations of actions and choices. 
This provides a complementary  verification of the aspects that are abstracted in the above algorithms. 
The implementation of {\tt CARE} has been formally modelled in {\sc Uppaal} as a network of timed automata. 
Figure~\ref{fig:runnableorchestration} depicts the automaton for the {\tt RunnableOrchestration}.
Due to lack of space, the automaton for the {\tt RunnableOrchestratedContract} and traceability information linking the model to the source code are available from~\cite{UppaalModels}.  
Both the synthesised orchestration (which is assumed to have been synthesised correctly) and other details specified in Algorithm~\ref{alg:orchestration} and Algorithm~\ref{alg:service} are abstracted away in the formal model.

The behaviour according to the given configuration  of action and choice is modelled inside each automaton.  
Global declarations include the number of services $N$, the size of the buffers, two variables {\tt action} and {\tt choice} storing the corresponding configuration for all automata, and the communication buffers. 
Java TCP/IP sockets communications are asynchronous with FIFO buffers.
In the model, arrays are used to encode these buffers that are only modified with functions for enqueueing and dequeuing messages.
Each party communicates with the partner using two buffers (one for sending and one for receiving). 
Both automata declare a method {\tt enqueue} for sending a message to the partner.  
Similarly, both automata have a method {\tt dequeue} for consuming messages from their respective buffers.  
According to the semantics of Java TCP/IP sockets, a transition having a send in its effect will check in its guard whether there is enough space left in the buffer of the partner by calling  either the method {\tt available} (returning the space left) or {\tt isFull}.
Moreover, before reading it is always checked whether the buffer is not empty with the method {\tt !isEmpty}. When the buffer is empty, the automaton blocks until a message is received.
The locations of the model are \emph{urgent} (denoted with {\tt U}) to guarantee that when the appropriate message is received it will eventually be consumed (i.e., there is no starvation). 
 
The absence of deadlocks was verified  by model checking the CTL formula 
\texttt{\scriptsize
A[\,](not deadlock || (ror.Terminated \&\& (forall(i:id\_t) ROC(i).Terminated))),}
in which {\tt ror} is the orchestrator and {\tt ROC(i)} is a runnable orchestrated contract identified with index~$i$.
Moreover, the absence of orphan messages was verified by model checking 
\texttt{\scriptsize
A[\,]((ror.Terminated \&\& (forall (i:id\_t) ROC(i).Terminated)) imply allEmpty()),}
in which the predicate {\scriptsize\tt allEmpty()} is satisfied when the buffers are empty. 
Finally,  \linebreak
\texttt{\scriptsize
A[\,](ror.Stop imply A<\,>(ror.Terminated \&\& (forall (i:id\_t) ROC(i).Terminated) \&\& allEmpty()))}
was used to verify that whenever a choice to stop is made, eventually all services and the orchestrator will terminate their execution. 

\vspace{-0.3cm}
\section{Building Applications with {\tt CARE}}\label{sect:methodology}
We now discuss how {\tt CARE} can be adopted to develop  applications with contract automata.
\begin{figure}[t]
    \centering
    \includegraphics[width=0.75\columnwidth]{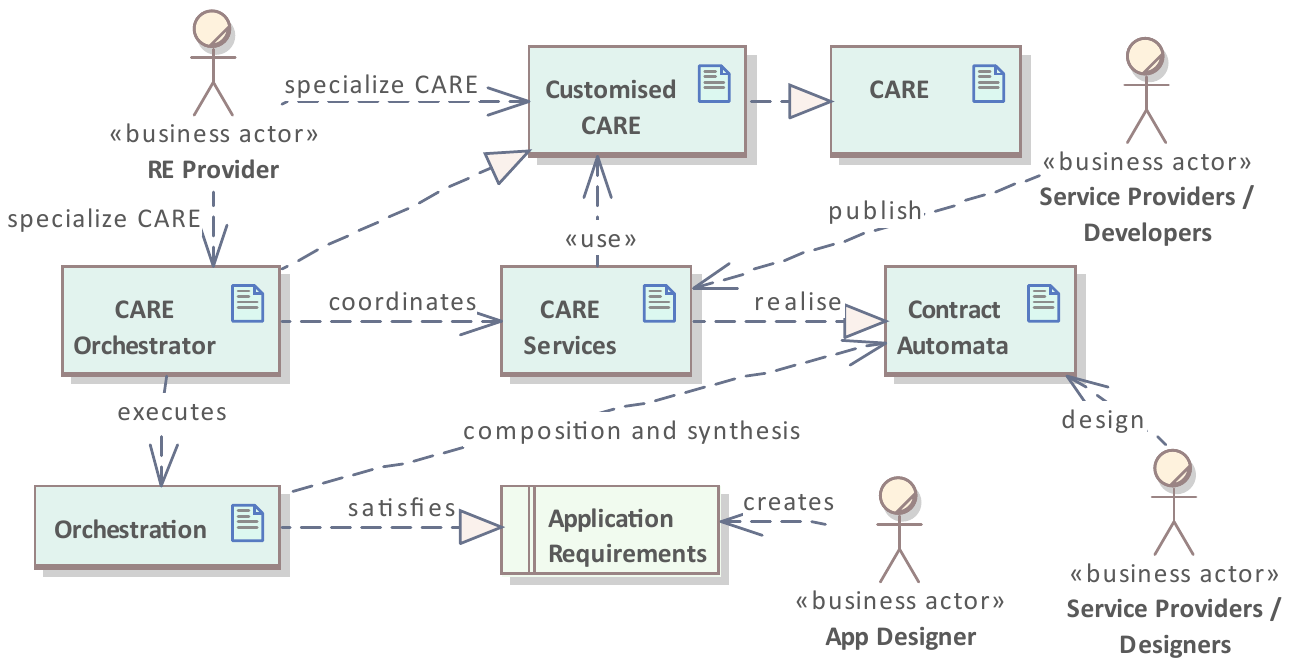}
    \vspace*{-0.25cm}
    \caption{\label{fig:methodology}The {\tt CARE} business actors developing contract-based applications}
\end{figure}
The diagram in Figure~\ref{fig:methodology} depicts the  responsibilities of the business actors involved in the overall realisation of contract-based applications using {\tt CARE}.
The first actor is the provider of the runtime environment ({\tt RE Provider} in Figure~\ref{fig:methodology}). 
This actor customises {\tt CARE} and its classes according to specific needs, possibly introducing new different options for choices and actions implementing the abstract methods provided by {\tt CARE} (described in Section~\ref{sect:design}), and delivers to the other actors a customised version of {\tt CARE}, which also comprehends an orchestrator. Note that this customisation is not necessary, but is a further possibility allowed by the   {\tt CARE} software design.

The second kind of actors are the service providers, who publish their contracts, implemented by remote (non-disclosed) Java classes, and use a \texttt{RunnableOrchestratedContract} to make their contract publicly accessible using {\tt CARE}, while hiding implementation details.
Service providers may choose among different realisations of their \texttt{RunnableOrchestratedContract}, provided by the first actor above. 
Notably, implementing each atomic action of a service and designing the interaction behaviour through contract automata are two different concerns. 
The designer (cf.\ Figure~\ref{fig:methodology}) specifying interactions as a contract is not required to be an expert in the underlying implementation technology (e.g., Java sockets), while 
the developer implementing actions and selecting the {\tt CARE} configuration is not required to be skilled in contract automata theory. 
The specification and implementation of a service can thus be seamlessly integrated using the facilities provided by {\tt CARE}.
This integration using {\tt CARE} is depicted with a {\em realize} arrow from the services to the contracts.  
Most importantly, when implementing the service, the developer does not need to worry about the underlying low-level interactions between  services, potential deadlocks and other communication issues.  
This error-prone implementation activity is already resolved  by {\tt CARE}, as discussed in Section~\ref{sect:correctness}.
This separation of concerns also solves the problem of ``muddling the main program logic with auxiliary logic related to error handling'' (i.e., handling the Java communication exceptions)~\cite{DBLP:series/lncs/FrancalanzaMT20}. 


The third actor is the application designer ({\tt App Designer} in Figure~\ref{fig:methodology}). 
This is a user of both the second and the first actor.  
The designer is responsible for specifying the  \emph{requirements} of the application, and to find a suitable set of remote services whose synthesised orchestration satisfies the desired requirements. 
Once the contracts are discovered, the orchestration enforcing the requirements is automatically synthesised as a new contract.  
This is depicted by an arrow from {\tt Orchestration} to {\tt Contract Automata} in Figure~\ref{fig:methodology}. 
The application designer exploits {\tt CARE}, choosing a specific implementation of \texttt{RunnableOrchestration} and {\tt RunnableOrchestratedContract}, passing as arguments the addresses of the services, as well as the synthesised  orchestration.
Formal results from contract automata theory~\cite{BBDLFGD20,BBP20,BDF16,BDFT15} (cf.\ Section~\ref{sect:background})  guarantee that no \texttt{ContractViolationException} will ever be raised at runtime (cf.\ Section~\ref{sect:correctness}).
Finally, note that one individual could take the roles of more actors if needed (e.g., covering both roles of developer and designer, designing a global requirement, implementing a new choice, and publishing a target contract). The proposed separation of concerns is logical. 
The roles and responsibilities of the various business actors described in this section are summarised in Table~\ref{tab:roles}.

\begin{table}[t]
    \caption{The roles and responsibilities of the business actors involved in developing applications specified via contract automata.}
    \label{tab:roles}
    \centering
    \scalebox{0.9}{
    \begin{tabular}{|p{4.5cm}||p{7.1cm}|}
    \hline
       {\bf Role}  & {\bf Responsibility}   \\ \hline\hline
        { \sf Runtime Environment Provider } &  Customisation of {\tt CARE}, implementation of abstract methods if needed \\ \hline
        { \sf Service Providers / Designers } &  Design contract automata and publish them \\ \hline
        { \sf Service Providers / Developers } & Implement the actions prescribed by contracts, select one of the available configurations of the runtime environment  \\ \hline
        { \sf App Designer } &  Design the requirements of the application, discover contracts, select one of the available configurations of the runtime environment    \\ \hline
    \end{tabular}
    }
\end{table}


\vspace*{-.25\baselineskip}
\section{Examples and Evaluation}\label{sect:usage}\label{sect:evaluation}
We discuss the usage of {\tt CARE}  using two examples.  
Their source code, video tutorials, and evaluation data are available from~\cite{CAREexample}.
%
%
%

\paragraph{\bf Alice and Bob}
This is a basic yet illustrative example. 
In this example, the requirement \texttt{req} of the application, designed by the {\tt App Designer},  is an automaton specifying  that an action \texttt{coffee} is observed after  an action \texttt{euro}.  
In this example, the {\tt RE Provider} will simply provide {\tt CARE} as it is, without further providing customised implementations of the abstract methods.

\begin{figure}[t]
    \centering
    \includegraphics[width=0.3\columnwidth]{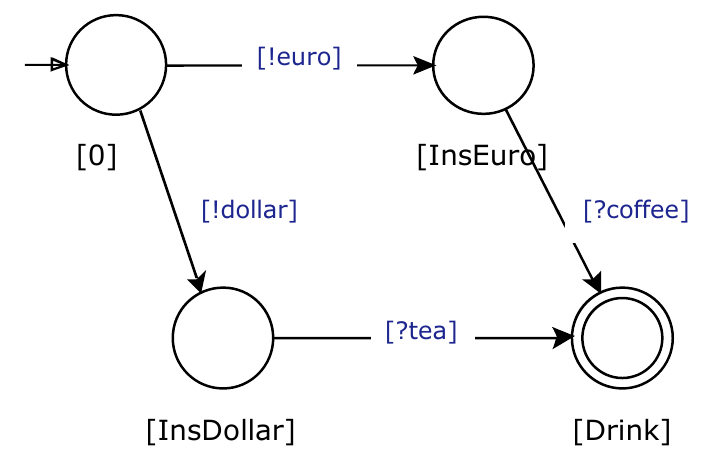} \hspace{-0.4cm}
    \includegraphics[width=0.3\columnwidth]{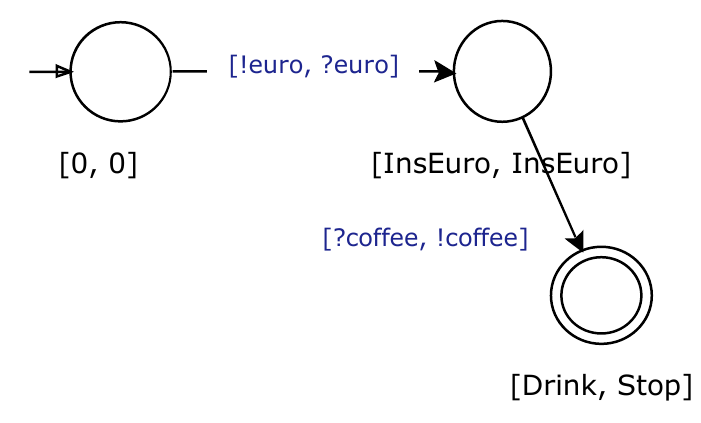}  \qquad
    \includegraphics[width=0.35\columnwidth]{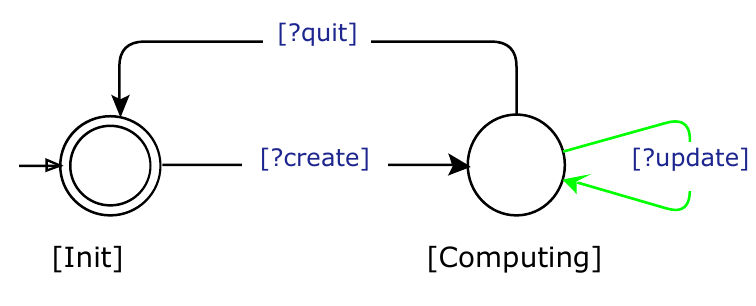}
    \caption{\label{fig:contracts}From left to right, the contract of \texttt{Alice}, the orchestration of {\tt Alice} and {\tt Bob} enforcing  the given requirement, and the contract of the {\tt Client}}
\end{figure}

We now move to the {\tt Service Provider / Designers} actors. Consider Figure~\ref{fig:contracts} (the automata have been constructed  using~\texttt{CAT\_App}).
The leftmost automaton is the contract of \texttt{Alice} and specifies that
\texttt{Alice} offers either a \texttt{!euro} or a \texttt{!dollar} to
her partner.
Then \texttt{Alice} requires \texttt{?coffee} or \texttt{?tea},
depending on which offer has been accepted.
Such a contract can be interpreted as describing the interaction pattern
of \texttt{Alice}, whilst abstracting away from the actual
implementation of each action.
To declare the signature of each contract action,
{\tt CARE}  uses Java Interfaces, as shown below.

\begin{minted}[mathescape,fontsize=\scriptsize]{java}
  public interface AliceInterface { 
    Integer coffee(String arg); Integer tea(String arg); 
    Integer euro(String arg); Integer dollar(String arg); }
\end{minted}
In the interpretation of contracts provided by {\tt CARE}, each contract
action is implemented by a method of an interface, whose names are in
correspondence. 
The implementation will be developed by the actor {\tt Service Provider / Developer}.
By implementing the corresponding interface it is possible to pair the
interaction logic described in Figure~\ref{fig:contracts} (left) with an
actual implementation, as shown below.

\begin{minted}[mathescape,fontsize=\scriptsize]{java}
  RunnableOrchestratedContract alice = new DictatorialChoiceRunnableOrchestratedContract(ca,
    8080,new Alice(),new CentralisedOrchestratedAction());
\end{minted}

The parameter {\tt ca} contains the leftmost contract in Figure~\ref{fig:contracts}.  The class \texttt{Alice} implements
\texttt{AliceInterface}.  This implementation is paired with the
corresponding contract:  the service listens to
port 8080 and the chosen implementation 
of the low level interactions is {\tt CentralisedOrchestratedAction}.
Notably, \texttt{RunnableOrchestratedContract} will take care of the low-level communications, abstracted away in \texttt{Alice.java}. 
In \texttt{AliceInterface}, each action
requires an argument (of type \texttt{String}) and returns a value (of
type \texttt{Integer}).  
During initialisation, each label of the contract  is extended with the 
information on the types of parameters and returned values from the interface, by 
instantiating a {\tt TypedCALabel}.
%
%
%
The contract of \texttt{Bob} is dual to the one of {\tt Alice} (i.e., all requests are turned to offers). 
The class {\tt RunnableOrchestration} can be instantiated as shown below.
\begin{minted}[mathescape, fontsize=\scriptsize]{java}
  RunnableOrchestration ror = new DictatorialChoiceRunnableOrchestration(req,new Agreement(),
    Arrays.asList(alice.getContract(),bob.getContract()),Arrays.asList(null,null), 
      Arrays.asList(alice.getPort(),bob.getPort()),new CentralisedOrchestratorAction());
\end{minted}
\texttt{DictatorialChoiceRunnableOrchestration} provides an implementation of the branch/termination selection where the orchestrator autonomously selects a branch. It is instantiated by passing as parameters  the requirement \texttt{req} to be enforced, the predicate on interactions among contracts (i.e., the property of agreement),  the list of contracts to compose, addresses and ports of the \texttt{RunnableOrchestratedContract} of \texttt{Alice} and \texttt{Bob}, and an object of class {\tt CentralisedOrchestratorAction}  implementing an {\tt OrchestratorAction}.
In this example, services are run locally on the same host as the orchestrator.
During instantiation, the contracts passed as arguments will be composed  to  synthesise their safe orchestration  in agreement.  

In this example, the contract of {\tt Bob} is in agreement with that of
{\tt Alice} (each request is matched by
a corresponding offer).
The orchestration {\tt orc} is the central automaton in Figure~\ref{fig:contracts}. 
After {\tt ror} has been instantiated, its  method {\tt isEmptyOrchestration()} is used to check if an agreement among 
contracts exists, i.e., if the synthesised orchestration is non-empty. 
During instantiation, {\tt RunnableOrchestration} also interacts with all services (using Java TCP sockets) to ensure that all share the same configuration, 
which in this case is a dictatorial choice with centralised action.  
If this is not the case, an exception is thrown. 
Upon successful instantiation, {\tt ror} can be executed to realise the application modelled through the requirement {\tt req} using the two contracts above.

Finally, we remark that it suffices to change the requirement to automatically adapt the services to generate a new application. In this example, if {\tt req} were changed to also allow a {\tt tea} in case of payment with {\tt dollar}, then  {\tt Alice} and {\tt Bob} could be adapted to fulfill this new requirement automatically. 

\paragraph{\bf Composition Service} 
Computing a composition of contract automata can be a costly operation.
For a front-end  running on a standard laptop (e.g., {\tt CAT\_App}), a desirable feature could be to delegate such costly computations to a remote service, hosted on a powerful machine.
This example showcases a service built with {\tt CARE} that computes a composition of contract automata.  
The service receives the operand automata together with other scalability options (e.g., a bound, invariants) from a client service.
The client service interacts through the console with a user who indicates which  automata to compose and the other options. 
{\tt CATLib} features on-the-fly bounded composition. When extending the bound of a previously computed  composition, the previously generated states of the composition are not recomputed.
The newly generated states are limited to those that exceeded the previous bound. 


The client contract is the rightmost automaton in Figure~\ref{fig:contracts}, whilst the service contract is dual (all requests are turned to offers).
The client contract can perform a necessary request {\tt update} from state {\tt Computing}.
This guarantees that in a non-empty orchestration, the necessary request of the client is matched by a corresponding offer.
If such a request were not necessary, a non-empty orchestration could also be obtained  when the client is composed with a service
that does not offer the {\tt update} action, but only actions {\tt create} and {\tt quit}.

From state {\tt Init}, the client can either terminate or 
perform a {\tt create} request. 
During the execution of this method the user  interacts  at console and types the needed input. The payload returned by the request method is submitted by the runtime support to the service executing the matching offer.   
The offer implementation takes as parameter the payload and returns  the composed automaton (which can be bounded to a specific depth), which is sent back to the requester. 
In the implementation of the {\tt update} request, the client sends an incremented bound to the service, which proceeds to compute the composition with the extended bound and sends it to the client.
The request {\tt quit} is used as a signal for resetting both the computed composition and the bound.

There are two choices: in state {\tt Init}, the orchestration can terminate or an action {\tt create} can be executed. 
In state {\tt Computing}, two possible actions can be performed. 
The {\tt MajoritarianChoiceRunnableOrchestratedContract} method\linebreak {\tt select}  is overridden by each service, to implement the specific choices to be made. 
The composition service always replies with an empty answer. 
This means that all choices are external to the service, the service does not indicate which choice has to be made.
The client service implements both choices as internal. The user of the client service will interact at console with the client service, and will indicate which choice has to be made. 
More details  can be found in~\cite{CAREexample}.

\paragraph{\bf Evaluation} 
We now measure the   advantages brought by adopting {\tt CARE}. 
To do so, we  compare two different implementations for the two examples. 
These two implementations of each example perform the exact same operations as described above.
Both implementations exploit the operations of composition and synthesis of contract automata provided by {\tt CATLib}. 
However, only one of them uses {\tt CARE} (as described above) whereas the other manually implements the prescribed  interactions between the services and and the orchestrator, without using any of the facilities provided by {\tt CARE}. 
In this way, it is possible to isolate and measure the benefits brought solely by using {\tt CARE}. 
These two implementations per example are open source and available for inspection from~\cite{CAREexample}, where they are located in two separate packages.

The comparison was performed using measures of code complexity as provided by {\tt SonarCloud}~\cite{DBLP:conf/icse/Campbell18}, an online service well integrated with {\tt GitHub} that performs, among others, continuous inspection of code quality and static analysis of code to detect bugs, and reports on code complexity.
We in particular used the code complexity reports feature of {\tt SonarCloud}. 
We used three different measures of complexity to showcase the benefits of using {\tt CARE}.
The first measure is the total amount of lines of code (thus excluding, among others, the lines of comments and white spaces).
We also adopted cyclomatic complexity~\cite{DBLP:journals/tse/McCabe76} and cognitive complexity~\cite{DBLP:conf/icse/Campbell18}. 
Cyclomatic complexity measures the number of independent paths in the software and it is a measure of code testability (this number is close to the number of branches to cover in the program). 
Cognitive complexity measures how difficult the control flow is to understand. This measure is roughly 
a counter incremented each time a control flow structure is encountered (e.g., {\tt if} and {\tt for}) and it is incremented commensurated with the level  of nesting of control flow structures (e.g., a first-level structure triggers an increment of~1, a second-level structure triggers an increment of~2, and so on)~\cite{DBLP:conf/icse/Campbell18}. 

\begin{table}[t]
    \caption{\label{tab:evaluation}Different measures of complexity of the examples from Section~\ref{sect:usage} implemented either with or without using {\tt CARE}}
    \vspace*{-0.25cm}
    \centering
    \scalebox{0.85}{
    \begin{tabular}{|l|l||r|r|r|}
    \hline
       \multicolumn{2}{|l||}{\bf } & {\sf LOC } & {\sf Cyclomatic Complexity } &  {\sf Cognitive Complexity } \\ \hline\hline
        \multirow{2}{*}{\bf Alice and Bob} & {\bf  without {\tt CARE }}  & 777 & 134  & 166 \\ \cline{2-5}
       & {\bf with {\tt CARE }} & 153 &  16 & 8 \\ \hline\hline
        \multirow{2}{*}{\bf Composition Service} & {\bf without {\tt CARE }}  & 854 & 155 & 211 \\ \cline{2-5}
        & {\bf with {\tt CARE }}  & 279 & 42 & 55  \\ \hline
    \end{tabular}
    }
\end{table}

The results are reported in Table~\ref{tab:evaluation}. 
To compare these quantities, we use the relative percent difference ($\textit{rpd}$): $\frac{|{\tt with CARE}-{\tt without CARE}|}{max({\tt with CARE},{\tt without CARE})}\times100$. This measures the change of complexity when using {\tt CARE} with respect to the reference value (i.e., {\tt withoutCARE}). 
The advantage of using {\tt CARE} is clear, 
as it drastically reduces the complexity of the software.
Indeed, when using {\tt CARE} (for the ``Alice and Bob'' and ``Composition Service'' examples, respectively) 
the $\textit{rpd}$ are: for the lines of code~$80.31\%$ and $67.33\%$, 
  for the cyclomatic complexity~$88.06\%$ and $72.90\%$, and
 for the cognitive complexity~$95.18\%$ and $73.93\%$.
%
%
These results are not surprising: the experiments  underline the complexity of the operations performed by {\tt CARE} and its key role in developing applications specified via contract automata. 
Indeed, for both examples the complexity of implementing the low-level communications is the dominant factor if compared to the interaction logic. 
This is more prominent for the ``Alice and Bob'' example, which in fact has greater $\textit{rpd}$ values. 
The burden of implementing these low-level communications among the services and the orchestrator is still on the developer side when not using {\tt CARE}.
We also remark how implementing the low-level communications is an error-prone activity that is completely delegated to the runtime support if one uses {\tt CARE}, thus improving the confidence in the correctness of the final application. 

\paragraph{\bf Scalability} 
The above experiments only measured the complexity of the software developed with or without {\tt CARE}.  
Another important aspect is the possibility of scaling to larger automata. 
{\tt CARE} is a runtime environment and does not face any scalability issue typical of static analysers (e.g., state-space explosion).   
On the other hand, the synthesis of a safe orchestration of contracts is computed using {\tt CATLib}, which may face scalability challenges when dealing with large compositions. 
In Section~\ref{sect:background},  the scalability features offered by {\tt CATLib} are reported. 
The performance of {\tt CATLib} has been previously measured in~\cite{BasileB21}. 
Concerning the formal verification of {\tt CARE} discussed in Section~\ref{sect:correctness}, we recall 
that the orchestration automaton is abstracted away in the {\sc Uppaal} model. 
Thus, the formal model of {\tt CARE} is verified for any orchestration of any size. 

On a side note, the single-responsibility principle~\cite{10.5555/515230}  advocates to assign a single responsibility to each class. By interpreting this principle over behavioural contracts, we conclude that a contract automaton assigned to a single class (e.g., the rightmost automaton in Figure~\ref{fig:contracts}) should not exhibit a large  behaviour.

\vspace*{-0.2cm}
\section{Conclusion}\label{sec:conclusion}

We have presented the first runtime environment for contract automata, called {\tt CARE}. 
Our proposal advances the state-of-the-art of the research on contract automata by showing a possible realisation of an orchestration engine, abstracted away in the contract automata theory, but needed for implementing applications specified with contract automata, and guaranteeing that the implementation of service-based  applications respect their specification.   
This contribution improves our understanding of the relation between a specification with contract automata and its implementation, and the corresponding level of abstraction.
%

With {\tt CARE}, it is  possible to promote a separation of concerns between formal methods experts specifying the expected behaviour using automata on one side, and developers (not required to be experts in formal methods) implementing the actions on the other.  
Furthermore, an application built using {\tt CARE} is based on rigorous theoretical results from the contract automata theory, guaranteeing properties such as absence of deadlocks and absence of orphan messages, reachability of final states, and absence of  \texttt{ContractViolationException}. 
Moreover, 
 {\tt CARE} promotes modularity of applications composed by different services that are reusable in different applications and that can be adapted to satisfy different requirements through the synthesis of well-behaving orchestrations.
Experiments showed the improvement in terms of decreased software complexity when using {\tt CARE} instead of manually implementing the low-level interactions among services implementing the operations prescribed by their contracts.

\paragraph{\bf Future Work}
\texttt{CATLib} already implements the  synthesis of choreographies~\cite{BBP20}, which {\tt CARE} will support in the future.
Although {\tt CARE} has been developed in the framework of contract automata, we plan to investigate the integration of this technology  with other behavioural types languages and tools (e.g., typestates). 





\paragraph{\bf Acknowledgment}
\begin{small}
Funded by MUR PRIN 2020TL3X8X project T-LADIES (Typeful Language Adaptation for Dynamic, Interacting and Evolving Systems).
\end{small}


\bibliographystyle{splncs04}
\bibliography{bib}

\end{document}